\documentclass[
  aps,
  prl,
  twocolumn,
  superscriptaddress,
  nofootinbib,
  longbibliography,
  floatfix
]{revtex4-2}

\usepackage{amssymb}
\usepackage{amsmath}
\usepackage{bm}
\usepackage{graphicx}
\usepackage[colorlinks=true,citecolor=blue,urlcolor=blue,linkcolor=blue]{hyperref}

\begin{document}

\title{Inclination Diffusion in Relativistic Loss Cones}

\author{Wenkang Xin}
\email{wenkang.xin@exeter.ox.ac.uk}
\affiliation{Department of Physics, University of Oxford, Parks Road, Oxford OX1 3PU, United Kingdom}

\date{\today}

\begin{abstract}
Relativistic capture and tidal disruption around a spinning black hole depend on both the magnitude and direction of the star's angular momentum,
yet loss-cone models often assume fixed orbital inclinations by ignoring the associated diffusion.
We show that this is not justified: for isotropic two-body relaxation near a small loss threshold,
angular-momentum magnitude $L$ and inclination $x = L_{z}/L$ diffuse on comparable timescales, $t_{E} \gg t_{L} \sim t_{x}$.
For Kerr capture, retaining inclination diffusion significantly amplifies the prograde--retrograde contrast while leaving the total inclination-integrated flux nearly unchanged.
An almost correct integrated flux can hide a badly wrong angular distribution.
The three-dimensional diffusion problem nevertheless retains enough angular structure to permit analytic treatment.
By representing pericenter removal as a continuous sink, we obtain a closed-form loss flux solution for a nearly linear Kerr tidal-disruption boundary, finding close agreement with phase-resolved calculations.
Inclination-dependent loss therefore requires inclination-resolved diffusion even when integrated rates appear robust.
\end{abstract}

\maketitle

\paragraph{Introduction.---}
Stellar encounters continually scatter stars toward low-angular-momentum orbits around massive black holes (BHs).
Stars that penetrate far enough may be swallowed by the BH's horizon or torn apart in a tidal disruption event (TDE).
The supply of stars into the dangerous region sets the rates of these events, which serve as an input to population studies of TDEs and extreme-mass-ratio-inspirals (EMRIs), and probes both BH demographics and the dynamics of galactic nuclei.

The orientation of orbits that induce such strong-field events affects the observable outcome.
For TDEs, spin-induced precession depends on the incoming orbital inclination and can alter debris-stream geometry and the subsequent circularization, thereby affecting the observable emission~\cite{Dai2013,Guillochon2015,Batra2023}.
Synchronized X-ray and radio modulation in TDEs has been modeled as Lense--Thirring coprecession of a misaligned disk--jet system, showing that orientation can leave a measurable imprint in the observed event light curve~\cite{Stone2012,Pasham2024,Wang2025a}.
For EMRIs, orbit inclination is an intrinsic gravitational-wave parameter measurable by the Laser Interferometer Space Antenna (LISA)~\cite{Barack2004}, and its population distribution may distinguish conventional ``dry'' EMRIs formed by two-body stellar encounters from the gas-assisted ``wet'' ones in active galactic nucleus (AGN) disks~\cite{Pan2021,Sun2026}.
An accurate total event rate of these processes can therefore still miss observable information carried by the inclination distribution.

Capture by the horizon and tidal disruption are different physical outcomes, but they enter stellar dynamics through the same object:
an angular-momentum threshold in the phase space of the stellar distribution function (DF), below which stars are killed within one orbital period.
Relaxation drives stars toward this loss threshold, whereas relativistic outcome physics fixes its position.

In this work, orientation refers to the general direction of the orbital angular-momentum vector $\mathbf{L}$ on the unit sphere, whereas inclination refers to its angle $\iota$ relative to the BH spin axis, represented by $x \equiv L_{z}/L = \cos{\iota}$.
In the axisymmetric Kerr problem, orientation dependence reduces to dependence on $x$.

\paragraph{Relativistic loss demands inclination dynamics.---}
Classical loss-cone theory describes how stellar encounters scatter stars to angular momenta small enough for capture or tidal disruption~\cite{Frank1976,Lightman1977,Cohn1978,Merritt2013a}.
In a spherical galaxy with a nonspinning Schwarzschild BH, the loss threshold depends only on energy $E$ and angular-momentum magnitude $L$, so the inclination $x \equiv L_{z}/L$ of angular momentum can be removed from the diffusion problem.
Kerr spin breaks this simplification: capture and tidal-disruption thresholds also depend on the inclination of the orbit relative to the spin~\cite{Kesden2012}, and the loss threshold becomes a surface $L_{\mathrm{lc}}(E, x)$.

A common way to accommodate an inclination-dependent boundary is to solve a family of loss-cone problems at fixed inclinations, a construction that has been used directly for EMRIs~\cite{AmaroSeoane2013,Zhang2026}.
Prior TDE calculations have bracketed the unresolved $x$-dependence using ``inclination-preserving'' (IP, zero $x$-diffusion) and ``isotropized'' (ISO, infinite $x$-diffusion) prescriptions~\cite{Singh2024}.
These are useful closures, but neither follows from the two-body dynamics they are intended to approximate:
the encounters that change the magnitude of angular momentum also rotate its direction.
As we derive, the relevant ordering of the diffusion timescales is
\begin{equation}
t_{E} \gg t_{L} \sim t_{x} \, .
\label{eq:timescale_ordering_heuristic}
\end{equation}
Near a small loss surface, energy diffusion is slow, while the magnitude and direction of the small angular-momentum vector evolve on comparable asymptotic timescales.
This ordering has a simple geometric interpretation:
a transverse velocity kick at large radius changes $\mathbf{L} = \mathbf{r} \times \mathbf{v}$;
the kick's components parallel and perpendicular to $\mathbf{L}$ respectively change its magnitude and direction, with neither effect parametrically suppressed.
The same kick produces only a small fractional change in orbital energy.

We test these consequences using two complementary relativistic loss processes around Kerr BHs.
The strongly inclination-dependent capture boundary provides a stringent test of whether inclination transport can be neglected, while the nearly linear tidal-disruption boundary isolates the first non-spherical mode and permits an analytic treatment.
The former establishes the necessity of inclination transport; the latter demonstrates the tractability of the additional dynamics.

\paragraph{Angular separation.---}
Consider the dynamical variables
\begin{equation}
E \, , \quad
Y \equiv L^{2} \, , \quad
x \equiv \frac{L_{z}}{L} \in [-1, 1] \, .
\label{eq:coordinates}
\end{equation}
For non-resonant two-body relaxation driven by an isotropic field-star distribution, transformation of the local diffusion tensor to the variables $(E, Y, x)$ gives
\begin{equation}
\bm{D}_{(E, Y, x)} =
\begin{pmatrix}
D_{EE} & D_{EY} & 0 \\
D_{EY} & D_{YY} & 0 \\
0 & 0 & D_{xx}
\end{pmatrix} \, , \quad
D_{Ex} = D_{Yx} = 0 \, ,
\label{eq:block_tensor}
\end{equation}
where the zeros are exact under the stated isotropy assumption.
The full derivation is given in End Matter and Ref.~\cite[Appx.~B~and~C]{Xin2026}.
Isotropy forces the stochastic increment in $x$ to have zero covariance with both $E$ and $Y$, although its diffusion rate still depends parametrically on $E$ and $Y$.
Related coefficients have been written in alternative coordinates~\cite{Vasiliev2013}; the ratio $L_{z}/L$ exposes the exact angular block and its spectral structure.

Near relativistic loss surfaces, rapid apsidal precession scrambles the argument of periapsis on a timescale short compared with the relaxation time (see Ref.~\cite[Ch.~4]{Xin2026} and End Matter).
After averaging over the apsidal angle, the drift and variance in $x$ combine so that the collision operator takes the form
\begin{equation}
\mathcal{L}_{\mathrm{coll}} = \mathcal{L}_{E, Y} + \mathcal{D}_{x}(E, Y) \mathcal{L}_{x} \, , \quad \mathcal{L}_{x} \equiv \partial_{x}\!\left[(1 - x^{2}) \partial_{x}\right] \, .
\label{eq:block_operator}
\end{equation}
Here $\mathcal{L}_{E, Y}$ contains the coupled energy--magnitude diffusion, while $\mathcal{D}_{x}(E, Y)$ retains its dependence on $E$ and $Y$.
The factor $1 - x^{2}$ enforces that the inclination flux vanishes at the poles $x = \pm 1$ so the distribution function (DF) remains in the physical domain.
It also has a geometric origin: $x$ represents the inclination of $\mathbf{L}$ on the unit sphere~\cite{Brown1963,Kocsis2015}, so isotropic rotational diffusion reduces to the axisymmetric Laplace--Beltrami operator $\nabla^{2}_{\Omega} = \partial_{x}[(1 - x^{2})\partial_{x}]$, whose regular eigenfunctions are the Legendre polynomials $P_{l}(x)$:
\begin{equation}
  \mathcal{L}_{x} P_{l}(x) = -l (l + 1) P_{l}(x) \, .
  \label{eq:legendre_eigenvalue}
\end{equation}

Inclusion of $x$ nevertheless preserves useful analytic structure.
Consider a general DF $f(E, Y, x, t)$, where $t$ is secular time.
Expanding the DF in Legendre polynomials $P_{l}(x)$ gives $f(E, Y, x, t) = \sum_{l = 0}^{\infty} f_{l}(E, Y, t) P_{l}(x)$, so Eqs.~\ref{eq:block_operator} and~\ref{eq:legendre_eigenvalue} give
\begin{equation}
\frac{\partial f_{l}}{\partial t} = \left[\mathcal{L}_{E, Y} - l (l + 1) \mathcal{D}_{x}(E, Y)\right] f_{l} \, .
\label{eq:legendre_hierarchy}
\end{equation}
Thus the three-dimensional collision problem becomes a hierarchy of two-dimensional equations in $(E, Y)$.
The bulk diffusion is therefore diagonal in $l$, while the geometry of the loss surface supplies the angular coupling.

With the bulk operator understood, the natural question is whether diffusion in each dimension operates on comparable timescales.

\paragraph{Magnitude and inclination relax together.---}
For a small loss surface, classical loss-cone theory gives the ordering $t_{Y}/t_{E} \sim C(E) Y / L_{c}^{2}(E) \ll 1$, where $L_{c}(E)$ is the circular-orbit angular momentum and $C(E)$ depends on the normalized orbit-averaged diffusion rates~\cite{Lightman1977,Cohn1978}.
Angular-momentum magnitude therefore relaxes at effectively fixed $E$, which can be treated as a parameter in the small-$Y$ limit.

Holding $E$ fixed and to leading order in $Y$, the angular-momentum sector of the local collision operator reads
\begin{equation}
\mathcal{L}_{Y, x}\rvert_{E} f \propto \left[\partial_{Y}\!\left(Y \partial_{Y} f\right) + \frac{1}{8 Y} \mathcal{L}_{x} f\right] \, ,
\label{eq:small_y_operator}
\end{equation}
where we drop the common prefactor for clarity (see Eq.~\ref{eq:local_small_y_operator} in End Matter).
For characteristic changes $\Delta Y/Y \sim 1$ and $\Delta x \sim 1$, the operator gives the timescales for order-unity changes:
\begin{equation}
t_{Y} \sim Y \, , \quad t_{x, l} \sim \frac{8 Y}{l (l + 1)} \, .
\label{eq:small_y_timescales}
\end{equation}
For every fixed low-$l$ mode, magnitude and inclination diffusion are of the same asymptotic order as $Y \to 0$.
Therefore, treating $x$ as a frozen label discards a leading-order transport process.

The above result provides the non-resonant baseline for inclination diffusion, which can be further enhanced by additional processes.
One channel is a non-spherical background potential, which can torque orbits and drive additional inclination transport~\cite{Einsel1999,Tep2022}.
The simple relations $D_{Ex} = D_{Yx} = 0$, and hence the separable angular operator derived here, need not survive.
Such departures do not, however, in general provide a justification for freezing $x$.

Another important example is vector resonant relaxation (VRR), which stochastically reorients orbital angular momenta while approximately conserving their magnitudes~\cite{Rauch1996}.
Kocsis and Tremaine~\cite{Kocsis2015} showed that VRR can be described as a random walk on the unit sphere; in the Brownian limit, this becomes spherical diffusion with spherical-harmonic modes.
For an axisymmetric problem, this reduces to the same Legendre operator $\mathcal{L}_{x}$ derived here, but with a different strength coefficient.

To compare different representations of inclination diffusion, we introduce a diagnostic multiplier $\Lambda$ such that $\mathcal{L}_{x} \to \Lambda \mathcal{L}_{x}$ in Eq.~\ref{eq:small_y_operator}.
$\Lambda = 1$ is the non-resonant two-body baseline, $\Lambda = 0$ is called ``independent slices'' (suppressing $x$-diffusion), and $\Lambda \to \infty$ is called ``rapid mixing'' (overwhelming $x$-diffusion).
$\Lambda \to \infty$ is not intended as a detailed model of VRR, but as the asymptote approached when additional diffusion dominates.

In general, Eq.~\ref{eq:small_y_timescales} shows that suppressing inclination diffusion is unjustified; whether doing so materially changes flux distributions depends on the geometry of the loss surface.
We first demonstrate this with Kerr capture, whose strong inclination dependence provides a deliberate stress test.

\paragraph{Phase-resolved loss and Kerr capture stress test.---}
Classical loss-cone theory models the loss process as a one-per-orbit removal of stars inside the loss surface, where the relevant dynamical variables are the slow orbital elements $(E, Y, x)$ (with $E$ fixed as a label) and the radial distance $r$.
In the phase-resolved picture, the DF is emptied at successive pericenter passages, leading to an explicit dependence on $r$. As is conventional in loss-cone literature, we parametrize the fast radial phase by a normalized variable $\tau \in [0, 1]$ that runs between successive pericenter passages (see, e.g., Refs.~\cite{Cohn1978,Merritt2013a,Xin2026}).

Consider the radial period $P$, which in general depends on $E$ and $Y$, but $P(E, Y) \approx P(E)$ in the small-$Y$ limit. Eq.~\ref{eq:small_y_operator} can be specialized to the phase-resolved Fokker--Planck equation
\begin{equation}
\frac{\partial f}{\partial \tau} = P(E) \bar{D}(E) \left[\partial_{R}\!\left(R \partial_{R} f\right) + \Lambda \partial_{x}\!\left(\frac{1 - x^{2}}{8 R} \partial_{x} f\right)\right] \, ,
\label{eq:2d_fp}
\end{equation}
where $\bar{D}(E)$ is an orbit-averaged diffusion coefficient and $R \equiv Y/L_{c}^{2}(E) \in [0, 1]$ is a normalized angular-momentum magnitude (see Eq.~\ref{eq:orbit_averaged_diffusion_rate} in End Matter).
The fullness parameter $q(E) \equiv P \bar{D} / R_{0}$ compares the dynamical timescale $P$ to the diffusion timescale $R_{0}/\bar{D}$, where $R_{0} \equiv \langle R_{\mathrm{lc}}(x) \rangle_{x}$ is the mean loss-surface size;
$q \ll 1$ and $q \gg 1$ are the empty- and full-cone limits, respectively~\cite{Merritt2013a}.
An $x$-dependent loss surface imposes the pericenter reset
\begin{equation}
f(R, x, 0^{+}) = \Theta[R - R_{\mathrm{lc}}(x)] f(R, x, 1^{-}) \, .
\label{eq:pericentre_reset}
\end{equation}
The DF inside $R < R_{\mathrm{lc}}(x)$ is removed once per orbit.
The loss rate is determined by the phase-space density removed at each pericenter passage; by number conservation, this equals the flux supplied to the loss region over the preceding orbit.
In the normalization used here, the inclination-differential loss is the DF removed immediately before pericenter:
\begin{equation}
\frac{\mathrm{d}F}{\mathrm{d}x}
\equiv \frac{1}{P \bar{D}} \int_{0}^{R_{\mathrm{lc}}(x)} f(R, x, 1^{-}) \, \mathrm{d}R \, .
\label{eq:phase_resolved_flux}
\end{equation}

We now use the strongly $x$-dependent Kerr capture surface as a stress test of setting $\Lambda = 0$ (independent slices) or $\Lambda \to \infty$ (rapid mixing) in Eq.~\ref{eq:2d_fp}.
At fixed relativistic energy $\mathcal{E}$, the separatrix between non-plunging and plunging geodesics defines the capture threshold $L_{\mathrm{cap}}(\mathcal{E}(E), x)$, conventionally referred to as the last stable orbit (LSO)~\cite{Stein2020}.
We adopt the marginally bound limit $\mathcal{E} = 1$, for which the relevant critical spherical orbit is the innermost bound spherical orbit (IBSO)~\cite{Wilkins1972,Hod2013,Mummery2023a}, thereby fixing the capture threshold $L_{\mathrm{cap}}(x)$.
The relativistic construction and boundary tables are described in Ref.~\cite{Xin2026}.
For dimensionless spin $a = 0.99$, the boundary used here satisfies
\begin{equation}
R_{\mathrm{cap}} = R_{0} \frac{L_{\mathrm{cap}}^{2}}{\langle L_{\mathrm{cap}}^{2} \rangle_{x}} \, , \quad \frac{\max{R_{\mathrm{cap}}}}{\min{R_{\mathrm{cap}}}} = 4.71 \, ,
\label{eq:lcap_boundary}
\end{equation}
where we adopt the representative mean size
$R_{0} \equiv \langle R_{\mathrm{cap}} \rangle_{x} = 10^{-5}$ for subsequent fixed-$E$ calculations.

\begin{figure}[!t]
\centering
\includegraphics[width=\columnwidth]{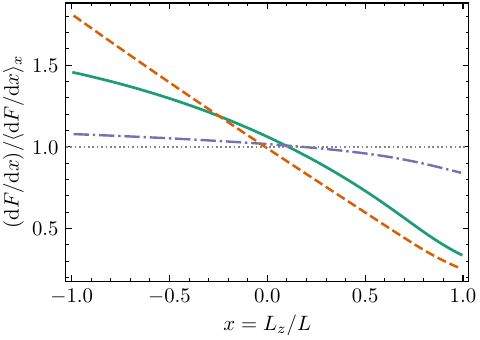}
\caption{
Kerr capture for $a = 0.99$.
Normalized angular capture flux at $q = 1$ under two-body relaxation $\Lambda = 1$ (green solid), independent slices $\Lambda = 0$ (purple dash-dotted), and rapid mixing $\Lambda \to \infty$ (orange dashed).
}
\label{fig:lcap_slices}
\end{figure}

We solve Eqs.~\ref{eq:2d_fp} and~\ref{eq:pericentre_reset} with $\Lambda = 0$, $\Lambda = 1$, and $\Lambda \to \infty$ over a range of $q$ values.
Numerical results are shown in Fig.~\ref{fig:lcap_slices}.
We define a baseline $x$-integrated flux $F_{\Lambda = 1} \equiv \int_{-1}^{1} (\mathrm{d}F_{\Lambda = 1}/\mathrm{d}x) \, \mathrm{d}x$ and an integrated error
\begin{equation}
\mathcal{E}_{\mathrm{ang}}
\equiv
\frac{1}{F_{\Lambda = 1}}
\int_{-1}^{1}
\left\lvert
\frac{\mathrm{d}F_{\Lambda}}{\mathrm{d}x}
- \frac{\mathrm{d}F_{\Lambda = 1}}{\mathrm{d}x}
\right\rvert \, \mathrm{d}x \, .
\label{eq:angular_error}
\end{equation}
We also use the following hemispheric contrast to quantify the redistribution of the angular flux:
\begin{equation}
F^{\pm} \equiv \pm \int_{0}^{\pm 1} \frac{\mathrm{d}F}{\mathrm{d}x} \, \mathrm{d}x \, , \quad
H \equiv \frac{F^{-} - F^{+}}{F^{-} + F^{+}} \in [-1, 1] \, ,
\label{eq:contrast_definition}
\end{equation}
where $H = 0$ corresponds to equal fluxes from the two hemispheres, while $H = \pm 1$ means that all losses occur from only one hemisphere.

For the independent-slice case $\Lambda = 0$, across $0.03 \leq q \leq 30$, $\mathcal{E}_{\mathrm{ang}}$ reaches $24\%$, while the total-flux error $|F_{\Lambda = 0}/F_{\Lambda = 1} - 1|$ remains below $3.7\%$.
For the rapid-mixing case $\Lambda \to \infty$, the angular error is $87\%$, while the total-flux error is below $2.4\%$.
At $q = 1$, the calculations give
\begin{equation}
\begin{split}
&H_{\Lambda = 0} = 0.051 \, , \quad H_{\Lambda = 1} = 0.285 \, , \quad H_{\Lambda \to \infty} = 0.397 \, , \\
&\frac{F_{\Lambda = 0}}{F_{\Lambda = 1}} - 1 = -1.4\% \, , \quad \frac{F_{\Lambda \to \infty}}{F_{\Lambda = 1}} - 1 = +0.8\% \, .
\end{split}
\end{equation}
Therefore, an almost correct integrated flux can conceal a seriously incorrect angular distribution if inclination diffusion is misrepresented as either suppressed or instantaneous.

\paragraph{Continuous loss and weakly dipolar tidal disruption.---}
Kerr capture establishes that inclination diffusion cannot in general be discarded, but its strong $x$-dependence makes analytical progress difficult.
We therefore turn to a weakly anisotropic test problem motivated by the Kerr tidal-disruption boundary.
For a solar-type star around a $10^{7} M_{\odot}$ Kerr BH with $a = 0.99$, the equatorial-pericenter prescription of Ref.~\cite{Xin2026} gives a dipole-dominated disruption boundary, with $c_{1} = -0.0797$ and leading non-dipolar coefficient $c_{2} = 0.0041$, where $R_{\mathrm{tde}}(x) \equiv L_{\mathrm{tde}}^{2}(x)/\langle L_{\mathrm{tde}}^{2} \rangle_{x} = 1 + \sum_{l} c_{l} P_{l}(x)$.
Since the physical boundary is nearly linear, we isolate its leading anisotropic mode and replace it by a pure-dipole boundary
\begin{equation}
R_{\mathrm{lc}}(x) = R_{0} [1 + \epsilon P_{1}(x)] \, , \quad P_{1}(x) = x \, , \quad |\epsilon| \ll 1 \, .
\label{eq:linear_boundary}
\end{equation}

Following the continuous-loss model introduced by Broggi~\cite{Broggi2025}, we replace once-per-orbit removal by a continuous sink term of the form
\begin{equation}
  \nu_{\mathrm{phys}} \equiv \frac{\Theta[R_{\mathrm{lc}} - R]}{P} \, .
\end{equation}
Using a dimensionless sink $\tilde{\nu}(R, x) \equiv \nu_{\mathrm{phys}}(R, x)/\bar{D}$, we extend Broggi's treatment to include the inclination dimension via the steady-state Fokker--Planck equation
\begin{equation}
\begin{split}
&0 = \partial_{R}\!\left(R \partial_{R} f\right)
+ \Lambda \partial_{x}\!\left(\frac{1 - x^{2}}{8 R} \partial_{x} f\right)
- \tilde{\nu}(R, x) f \, , \\
&\tilde{\nu}(R, x) = \frac{\Theta[R_{\mathrm{lc}}(x) - R]}{q R_{0}} \, , \quad q \equiv \frac{P \bar{D}}{R_{0}} \, ,
\end{split}
\label{eq:tde_reaction_diffusion}
\end{equation}
where $\Lambda$ controls the diffusion strength and the fullness $q$ has the same definition as in the phase-resolved problem.
Note that the sink term removes the explicit dependence on the radial phase $\tau$ and the pericenter reset condition.
Expand in $\epsilon$ and isolate the first non-spherical response:
\begin{equation}
\begin{split}
\tilde{\nu}(R, x)
&= \frac{\Theta(R_{0} - R)}{q R_{0}}
+ \epsilon \frac{\delta(R - R_{0})}{q} P_{1}(x)
+ \mathcal{O}(\epsilon^{2}) \, ,\\
f(R, x)
&= f^{(0)}(R) + \epsilon g(R) P_{1}(x)
+ \mathcal{O}(\epsilon^{2}) \, .
\end{split}
\label{eq:dipole_ansatz}
\end{equation}
Because the unperturbed case is the monopole, the $P_{1}$ deformation sources only $l = 1$.
This problem has an analytic solution presented in End Matter.
The differential loss is analogous to the phase-resolved version in Eq.~\ref{eq:phase_resolved_flux}:
\begin{equation}
\frac{\mathrm{d}F^{\mathrm{sink}}_{\Lambda}}{\mathrm{d}x}
= \int \tilde{\nu} f \, \mathrm{d}R
= \frac{1}{q R_{0}} \int_{0}^{R_{\mathrm{lc}}(x)} f \, \mathrm{d}R \, ,
\label{eq:sink_flux_integral}
\end{equation}
where we define a general notation for either model:
\begin{equation}
\begin{split}
&\frac{\mathrm{d}F^{X}_{\Lambda}}{\mathrm{d}x}
= {\left(\frac{\mathrm{d}F^{X}_{\Lambda}}{\mathrm{d}x}\right)}_{\mathrm{iso}}
\left[1 + \epsilon A^{X}_{\Lambda} P_{1}(x)
+ \mathcal{O}(\epsilon^{2})\right] \, , \\
&X \in \{\mathrm{phase}, \mathrm{sink}\} \, .
\end{split}
\label{eq:general_flux_response}
\end{equation}
We stress that the continuous sink is not microscopically identical to phase-resolved removal.
By averaging over the radial phase, the discrete pericenter reset is replaced by a continuous drain with rate $1/P$.
Effectively, the sink term attaches a Poisson destruction clock with mean waiting time $P$ to every star inside the loss region.
While both prescriptions share the same mean loss rate, they differ most in the intermediate regime $q \sim 1$~\cite{Broggi2025}, where the dynamical and diffusion timescales are comparable.

Under the baseline two-body relaxation $\Lambda = 1$, the first-order response $A^{\mathrm{sink}}_{\Lambda = 1}$ can be evaluated in closed form.
Its limits are
\begin{equation}
\lim_{q \to 0} A^{\mathrm{sink}}_{\Lambda = 1} = \frac{1 + R_{0}}{2 (1 - R_{0})} \simeq \frac{1}{2} \, , \quad \lim_{q \to \infty} A^{\mathrm{sink}}_{\Lambda = 1} = 1 \, .
\label{eq:A_sink_limits}
\end{equation}
These limits of the first-order response exactly match those of the phase-resolved problem (see Eqs.~\ref{eq:A_lambda_phase_empty} and~\ref{eq:A_lambda_sink_empty_endmatter} in End Matter and Ref.~\cite{Xin2026}), which is a nontrivial validation of the continuous-loss approximation.
Because $P_{1}$ integrates to zero, the total flux only changes at second order in $\epsilon$.
This is expected since $\epsilon \to -\epsilon$ is equivalent to a reflection $x \to -x$.
This amounts to flipping the direction of the BH's spin in an isotropic stellar distribution, which cannot change the total flux.

\begin{figure}[!t]
\centering
\includegraphics[width=\columnwidth]{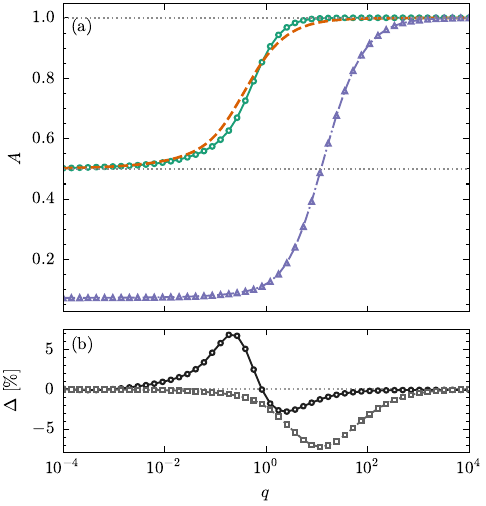}
\caption{
Dipole response at a fixed $R_{0} = 10^{-6}$ mean boundary.
(a) Phase-resolved two-body baseline $A^{\mathrm{phase}}_{\Lambda = 1}$ (green solid circles),
continuous-loss $A^{\mathrm{sink}}_{\Lambda = 1}$ (orange dashed),
and phase-resolved independent-slice $A^{\mathrm{phase}}_{\Lambda = 0}$ (purple dash-dotted triangles);
dotted lines mark analytical endpoint limits.
(b) The dipole residual $\Delta_{A} = A^{\mathrm{sink}}_{\Lambda = 1}/A^{\mathrm{phase}}_{\Lambda = 1} - 1$ (dark-grey circles), whose maximum absolute value is $7\%$, and the monopole residual $\Delta_{0} = (\mathrm{d}F^{\mathrm{sink}}_{\Lambda = 1}/\mathrm{d}x)/(\mathrm{d}F^{\mathrm{phase}}_{\Lambda = 1}/\mathrm{d}x) - 1$ (light-grey squares), whose maximum absolute value is $7.2\%$.
}
\label{fig:response_calibration}
\end{figure}

Fig.~\ref{fig:response_calibration} shows the continuous-loss response $A^{\mathrm{sink}}_{\Lambda = 1}$ together with the phase-resolved two-body baseline $A^{\mathrm{phase}}_{\Lambda = 1}$ and phase-resolved independent-slice $A^{\mathrm{phase}}_{\Lambda = 0}$.
The continuous-loss model captures the weak-dipole response to within about $6.9\%$ across loss-cone regimes.
By contrast, the independent-slice calculation gives a much smaller response and converges to a different empty-cone limit.
Indeed, this limit can be derived analytically to be $1/\ln{(1/R_{0})}$ (see Eq.~\ref{eq:A_phase_empty_special} in End Matter), which is qualitatively different from the expected baseline limit $1/2$.
On the other hand, for the monopole $(\mathrm{d}F^{X}_{\Lambda}/\mathrm{d}x)_{\mathrm{iso}}$ that represents the total flux,
the independent-slice calculation agrees with the two-body baseline to machine precision across $q$ because they share the same analytical expression, while the continuous loss differs by less than $7.2\%$.

The TDE calculation therefore repeats the lesson of the Kerr stress test: suppressing diffusion in $x$ causes a large error in the inclination-differential flux, even when the integrated flux is almost correct.

\paragraph{General structure of relativistic loss.---}
The fixed-$E$ calculations above provide concrete examples of why inclination transport cannot be neglected and how the additional dimension can nevertheless remain tractable.
They are particular realizations of the angular structure exposed by Eq.~\ref{eq:block_operator}.
We now return to the general three-dimensional problem: the bulk collisional dynamics remains diagonal in Legendre modes, while the inclination dependence of the relativistic loss surface supplies the mode coupling.

For deterministic pericenter removal, introduce the phase-resolved DF $f(E, Y, x, \tau; t)$, where $\tau \in [0, 1]$ is the normalized radial phase between successive pericenter passages.
At fixed secular time, the loss surface imposes
\begin{equation}
f(E, Y, x, 0^{+}; t) = \Theta[Y - Y_{\mathrm{lc}}(E, x)] f(E, Y, x, 1^{-}; t) \, .
\end{equation}
Expanding in Legendre polynomials gives
\begin{equation}
\begin{split}
f_{l}(0^{+})
&=
\sum_{m = 0}^{\infty}
S_{lm}(E, Y) f_{m}(1^{-}) \, ,
\\
S_{lm}
&\equiv
\frac{2 l + 1}{2}
\int_{-1}^{1}
P_{l}(x)
\Theta\!\left[Y - Y_{\mathrm{lc}}(E, x)\right]
P_{m}(x) \, \mathrm{d}x \, .
\end{split}
\label{eq:reset_hierarchy}
\end{equation}
The modes evolve independently between pericenter passages, while the loss surface couples them at the reset. The explicit radial-phase dependence nevertheless makes the intermediate-regime problem difficult.

The continuous loss removes the radial phase coordinate.
For the sink
\begin{equation}
\nu(E, Y, x) = \frac{\Theta[Y_{\mathrm{lc}}(E, x) - Y]}{P(E, Y)} \, ,
\end{equation}
projecting the Fokker--Planck equation gives
\begin{equation}
\begin{split}
\frac{\partial f_{l}}{\partial t}
&=
\left[
\mathcal{L}_{E, Y}
- l (l + 1) \mathcal{D}_{x}(E, Y)
\right] f_{l}
-
\sum_{m = 0}^{\infty}
\nu_{lm}(E, Y) f_{m} \, ,
\\
\nu_{lm}
&\equiv
\frac{2 l + 1}{2}
\int_{-1}^{1}
P_{l}(x) \nu(E, Y, x) P_{m}(x) \, \mathrm{d}x \, .
\end{split}
\label{eq:reaction_hierarchy}
\end{equation}
This is an autonomous hierarchy of coupled two-dimensional $(E, Y)$ equations.
The Legendre reduction is therefore a property of the collisional dynamics, independent of how loss is implemented;
the loss prescription determines only how the modes are coupled.
The weak-dipole calculation above isolates the first non-spherical member of this family and shows explicitly that restoring $x$ as a dynamical coordinate need not make relativistic loss-cone theory intractable.

\paragraph{Conclusion.---}
Black-hole spin makes relativistic loss processes inclination-dependent.
The small-loss-cone approximation freezes energy, not angular momentum or inclination.
For isotropic two-body relaxation, the ordering $t_{E} \gg t_{Y} \sim t_{x}$ means that treating $x$ as a passive label discards a leading-order diffusion process.
The Kerr capture calculation makes the consequence explicit:
at intermediate loss-cone fullness, misrepresenting diffusion in $x$ strongly alters the prograde--retrograde contrast while leaving the integrated flux nearly unchanged.
Neither independent-slice nor rapid-mixing approximations capture the correct angular distribution, even when the total flux is roughly correct.

Nevertheless, the variable $x = L_{z}/L$ exposes an exact diffusion block separating $x$ from $(E, Y)$; after apsidal averaging, the angular diffusion is diagonal in Legendre modes, with $x$-dependent loss supplying the mode coupling.
The weak-dipole TDE calculation solves the first non-spherical member of this hierarchy analytically, agreeing with phase-resolved removal at the ${\sim}7\%$ level over the tested range.

Orbital orientation should therefore not be treated as a frozen parameter or as a rapidly mixed quantity, but as a dynamical variable.
Orientation-dependent relativistic observables require orientation-resolved transport, even when the integrated loss fluxes appear robust.

\paragraph{Acknowledgments.---}
The author thanks James Binney for supervising the dissertation from which this work originated.

\bibliography{tidal_rate}

\newpage
\appendix

\section{Diffusion block}

We use the velocity coordinates $(v, \mu \equiv \cos{\theta_{v}}, \phi_{v})$ and transform to
\begin{equation}
E = \frac{v^{2}}{2} + \Phi(r) \, , \quad
Y = r^{2} v^{2} (1 - \mu^{2}) \, , \quad
x = \sin{\theta_{r}} \sin{\phi_{v}} \, ,
\label{eq:EYx_coordinates_diffusion_block}
\end{equation}
where $\Phi(r)$ is the gravitational potential, and $\theta_{r}$ is the polar angle of the position vector $\mathbf{r}$ and $\phi_{v}$ is the azimuthal angle of the velocity vector $\mathbf{v}$.
For an isotropic field-star distribution, write
\begin{equation}
T^{i} = \frac{v^{i}}{v} A_{\parallel} \, , \quad S^{ij} = \frac{v^{i} v^{j}}{v^{2}} B_{\parallel}
+ \left(\delta^{ij} - \frac{v^{i} v^{j}}{v^{2}}\right) \frac{B_{\perp}}{2} \, ,
\label{eq:isotropic_velocity_moments_diffusion_block}
\end{equation}
where $A_{\parallel} = \langle\Delta v_{\parallel}\rangle$, $B_{\parallel} = \langle(\Delta v_{\parallel})^{2}\rangle$, and $B_{\perp} = \langle(\Delta v_{\perp})^{2}\rangle$.
The standard coordinate transformation (see, e.g., Ref.~\cite[Ch.~5]{Merritt2013a}) gives
\begin{equation}
\begin{split}
\langle\Delta V^{\lambda}\rangle
&= \partial_{i} V^{\lambda} T^{i}
+ \frac{1}{2}\left(\partial_{i}\partial_{j} V^{\lambda}
- \Gamma^{k}_{ij} \partial_{k} V^{\lambda}\right) S^{ij} \, , \\
\langle\Delta V^{\lambda} \Delta V^{\nu}\rangle
&= \partial_{i} V^{\lambda} \partial_{j} V^{\nu} S^{ij} \, ,
\end{split}
\label{eq:coordinate_transform_diffusion_block}
\end{equation}
which gives
\begin{equation}
\begin{split}
\langle\Delta E\rangle
&= v A_{\parallel} + \frac{B_{\parallel} + B_{\perp}}{2} \, , \\
\langle\Delta Y\rangle
&= \frac{2 Y}{v} A_{\parallel} + \frac{Y}{v^{2}} B_{\parallel}
+ r^{2} B_{\perp} - \frac{Y}{2 v^{2}} B_{\perp} \, , \\
\langle\Delta x\rangle
&= -\frac{r^{2} B_{\perp}}{4 Y} x \, ,
\end{split}
\label{eq:full_drifts_diffusion_block}
\end{equation}
and the nonzero second moments
\begin{equation}
\begin{split}
&\langle(\Delta E)^{2}\rangle
= v^{2} B_{\parallel} \, , \quad \langle\Delta E \, \Delta Y\rangle
= 2 Y B_{\parallel} \, , \\
&\langle(\Delta Y)^{2}\rangle
= \frac{4 Y^{2}}{v^{2}} B_{\parallel} + 2 r^{2} Y B_{\perp}
- \frac{2 Y^{2}}{v^{2}} B_{\perp} \, , \\
&\langle(\Delta x)^{2}\rangle
= \frac{r^{2} B_{\perp}}{2 Y} (\sin^{2}{\theta_{r}} - x^{2}) \, ,
\end{split}
\label{eq:full_diffusion_block}
\end{equation}
$x$ has no mixed diffusion with either $E$ or $Y$, and this does not require the highly eccentric limit~\cite{Xin2026}.
Note that $x = \cos{\iota}$ and $\cos{\theta_{r}} = \sin{\iota} \cos{(\varpi + \psi)}$, where $\iota$ is the inclination angle, $\varpi$ the apsidal angle, and $\psi$ the true anomaly.
Averaging over the fast apsidal angle gives $\left\langle\sin^{2}{(\varpi + \psi)}\right\rangle_{\varpi} = 1/2$ and $\left\langle\sin^{2}{\theta_{r}}\right\rangle_{\varpi} = (1 + x^{2})/2$.
Substituting these averages into the general $x$-coefficients in Eqs.~\ref{eq:full_drifts_diffusion_block} and~\ref{eq:full_diffusion_block} yields the desired Legendre form in Eq.~\ref{eq:block_operator}.
We work in the standard regime in which apsidal phase mixing is fast compared with the collisional evolution of the distribution function.
This regime is generally appropriate in near-Keplerian galactic nuclei owing to relativistic and stellar-mass precession~\cite{Rauch1996,Kocsis2015,Sridhar2016}.

At fixed $E$, set $R = Y/L_{c}^{2}(E)$ and define the local and orbit-averaged diffusion rates
\begin{equation}
D(E, r) \equiv \frac{r^{2} B_{\perp}}{L_{c}^{2}(E)} \, , \quad
\bar{D}(E) \equiv \frac{1}{P(E)}
\oint \frac{\mathrm{d}r}{v_{r}} D(E, r) \, .
\label{eq:orbit_averaged_diffusion_rate}
\end{equation}
To leading order in $R \ll 1$,
\begin{equation}
\begin{split}
&\langle\Delta R\rangle_{E} = D + \mathcal{O}(R) \, , \quad
\langle(\Delta R)^{2}\rangle_{E} = 2 D R + \mathcal{O}(R^{2}) \, , \\
&\langle\Delta x\rangle_{E} = -\frac{D}{4} \frac{x}{R} \, , \quad
\langle\Delta R \, \Delta x\rangle_{E} = 0 \, , \\
&\langle(\Delta x)^{2}\rangle_{E}
= \frac{D}{2 R} (\sin^{2}{\theta_{r}} - x^{2}) = \frac{D}{4 R} (1 - x^{2}) \, ,
\end{split}
\label{eq:fixed_E_small_R_coefficients}
\end{equation}
where the last line follows from the apsidal average.

\section{Phase-resolved loss and dipole response}

Before orbit averaging, the local small-$Y$ collision operator at fixed $E$ obtained from Eq.~\ref{eq:fixed_E_small_R_coefficients} is
\begin{equation}
\mathcal{L}_{\mathrm{coll}} f
= D(E, r) \left[
\partial_{R}\!\left(R \partial_{R} f\right)
+ \partial_{x}\!\left(\frac{1 - x^{2}}{8 R} \partial_{x} f\right)
\right] \, .
\label{eq:local_small_y_operator}
\end{equation}
The Boltzmann equation retains its dependence on the local radial coordinate $r$:
\begin{equation}
\frac{\mathrm{d}f}{\mathrm{d}t}
= \frac{\partial f}{\partial t} + v_{r} \frac{\partial f}{\partial r}
= \mathcal{L}_{\mathrm{coll}} f \, .
\label{eq:local_boltzmann_streaming}
\end{equation}
Following Ref.~\cite{Cohn1978}, we define the normalized radial phase
\begin{equation}
\tau(E, R) \equiv \frac{1}{\bar{D}(E) P(E)} \int_{r_{-}}^{r} \frac{\mathrm{d}r}{v_{r}} D \, ,
\label{eq:tau_definition}
\end{equation}
where $r_{-}$ and $r_{+}$ are the pericenter and apocenter of the orbit, respectively, and $D(E, r)$ and $\bar{D}(E)$ are defined in Eq.~\ref{eq:orbit_averaged_diffusion_rate}.
Thus $\tau$ increases from $0$ to $1$ over one radial period and satisfies $v_{r} \partial_{r} \tau = D/(P \bar{D})$.
In steady state, $\partial f/\partial t = 0$ but the streaming term remains; substituting Eq.~\ref{eq:local_small_y_operator} into Eq.~\ref{eq:local_boltzmann_streaming} and changing from $r$ to $\tau$ gives Eq.~\ref{eq:2d_fp}.

For general $q$, the phase-resolved boundary-layer problem in Eq.~\ref{eq:2d_fp} has no known closed-form solution.
The empty- and full-cone limits, however, can be solved analytically.
Consider the weakly dipolar loss surface $R_{\mathrm{lc}}(x) = R_{0} [1 + \epsilon P_{1}(x)]$ with $P_{1}(x) = x$.
Outside the loss region, we ignore any $\tau$-dependence and write
\begin{equation}
f(R, x)
=
f_{1} + F_{0} \ln{R}
+ \epsilon g(R) P_{1}(x)
+ \mathcal{O}(\epsilon^{2}) \, ,
\end{equation}
where $F_{0}$ is the monopole flux, $g(R)$ is the dipole response, and $f_{1} = f(R = 1)$ is the outer boundary value where the DF is pinned to the spherical background and has no $x$-dependence~\cite{Xin2026}.
Using $\mathcal{L}_{x} P_{1} = -2 P_{1}$ and Eq.~\ref{eq:2d_fp} with $\partial f/\partial\tau = 0$, we have
\begin{equation}
R g'' + g' - \frac{\Lambda}{4 R} g = 0 \, .
\label{eq:phase_empty_dipole}
\end{equation}
With $\alpha \equiv \sqrt{\Lambda}$ and $g(1) = 0$,
\begin{equation}
g(R)
=
C_{\alpha}
\left(
R^{\alpha/2} - R^{-\alpha/2}
\right) \, .
\end{equation}
In the empty-cone limit $q \to 0$, depletion is much faster than diffusion, so the DF drops to zero at the loss boundary.
Expanding the absorbing condition $f[R_{\mathrm{lc}}(x), x] = 0$ about $R_{0}$ gives $g(R_{0}) = -F_{0}$.
Solving for the corresponding flux gives the first-order response
\begin{equation}
\lim_{q \to 0} A_{\Lambda}^{\mathrm{phase}}
=
\frac{\sqrt{\Lambda}}{2}
\coth{\left[
\frac{\sqrt{\Lambda}}{2}
\ln{\left(\frac{1}{R_{0}}\right)}
\right]} \, .
\label{eq:A_lambda_phase_empty}
\end{equation}
In particular,
\begin{equation}
\lim_{q \to 0} A_{\Lambda = 0}^{\mathrm{phase}}
=
\frac{1}{\ln{(1/R_{0})}} \, ,
\quad
\lim_{q \to 0} A_{\Lambda = 1}^{\mathrm{phase}}
=
\frac{1 + R_{0}}{2 (1 - R_{0})} \, .
\label{eq:A_phase_empty_special}
\end{equation}

For $q \to \infty$, diffusion is much faster than depletion, so the DF is uniform across the loss region.
The local flux is simply proportional to the local phase volume $R_{\mathrm{lc}}(x)$ removed per orbit:
\begin{equation}
\frac{\mathrm{d}F_{\Lambda}^{\mathrm{phase}}}{\mathrm{d}x}
\propto
R_{\mathrm{lc}}(x)
=
R_{0} [1 + \epsilon P_{1}(x)] \, ,
\end{equation}
and therefore
\begin{equation}
\lim_{q \to \infty} A_{\Lambda}^{\mathrm{phase}} = 1 \, .
\label{eq:A_lambda_phase_full}
\end{equation}
These are also the endpoint limits of the continuous-loss problem derived below; the two prescriptions differ only at intermediate loss-cone fullness.

\section{Dipole matching for the continuous loss}

We generalize the calculation in Ref.~\cite{Broggi2025} to include $x$-diffusion and an arbitrary diffusion strength $\Lambda$.
Substitution of Eq.~\ref{eq:dipole_ansatz} into Eq.~\ref{eq:tde_reaction_diffusion} with $\mathcal{L}_{x} P_{1} = -2 P_{1}$ gives
\begin{equation}
R g'' + g'
- \left[
\frac{\Theta(R_{0} - R)}{q R_{0}}
+ \frac{\Lambda}{4 R}
\right] g
=
\frac{f^{(0)}(R_{0})}{q} \delta(R - R_{0}) \, .
\label{eq:dipole_equation_endmatter}
\end{equation}
The monopole is the same as reported in Ref.~\cite{Broggi2025}:
\begin{equation}
f^{(0)}(R) =
\begin{cases}
B_{0} I_{0}\!\left(2\sqrt{R/(q R_{0})}\right) \, , & R < R_{0} \, , \\
f_{1} + F_{0} \ln{R} \, , & R > R_{0} \, ,
\end{cases}
\end{equation}
where $I_{j}$ is the modified Bessel function of the first kind of order $j$ and
\begin{equation}
F_{0} = \frac{f_{1}}
{\sqrt{q} I_{0}(z_{0})/I_{1}(z_{0}) + \ln{(1/R_{0})}} \, ,
\quad
z_{0} \equiv \frac{2}{\sqrt{q}} \, .
\end{equation}
It is useful to define
\begin{equation}
h(q) \equiv
\frac{I_{0}(z_{0})}{\sqrt{q} \, I_{1}(z_{0})} \, ,
\quad
\frac{f^{(0)}(R_{0})}{q} = F_{0} h(q) \, .
\label{eq:h_definition_endmatter}
\end{equation}
Let $\alpha \equiv \sqrt{\Lambda}$.
The regular interior solution and the exterior solution satisfying $g(1) = 0$ are
\begin{equation}
g(R) =
\begin{cases}
B_{\alpha} I_{\alpha}\!\left(2\sqrt{R/(q R_{0})}\right) \, , & R < R_{0} \, , \\
C_{\alpha} \left(R^{\alpha/2} - R^{-\alpha/2}\right) \, , & R > R_{0} \, .
\end{cases}
\label{eq:g_general_lambda_endmatter}
\end{equation}
Define
\begin{equation}
\begin{split}
\Phi_{\alpha}^{(0)}
&\equiv R_{0}^{\alpha/2} - R_{0}^{-\alpha/2} \, , \\
\Phi_{\alpha}^{(1)}
&\equiv \frac{\alpha}{2}
\left(R_{0}^{\alpha/2} + R_{0}^{-\alpha/2}\right) \, , \\
k_{\alpha}(q)
&\equiv
\frac{I_{\alpha}'(z_{0})}
{\sqrt{q} \, I_{\alpha}(z_{0})} \, , \\
\mathcal{J}_{\alpha}(q)
&\equiv
\frac{1}{2 I_{\alpha}(z_{0})}
\int_{0}^{z_{0}} u I_{\alpha}(u) \, \mathrm{d}u \, .
\end{split}
\label{eq:response_functions_endmatter}
\end{equation}
Continuity of $g$ and integration of Eq.~\ref{eq:dipole_equation_endmatter} across $R_{0}$ give
\begin{equation}
C_{\alpha} = F_{0} \, \frac{h(q)}{\Phi_{\alpha}^{(1)} - k_{\alpha}(q) \Phi_{\alpha}^{(0)}} \, .
\label{eq:C_alpha_endmatter}
\end{equation}
The loss flux is
\begin{equation}
\frac{\mathrm{d}F_{\Lambda}^{\mathrm{sink}}}{\mathrm{d}x} = \frac{1}{q R_{0}} \int_{0}^{R_{\mathrm{lc}}(x)} f(R, x) \, \mathrm{d}R \, ,
\end{equation}
and expanding to first order in $\epsilon$ gives
\begin{equation}
\frac{\mathrm{d}F_{\Lambda}^{\mathrm{sink}}}{\mathrm{d}x} = F_{0} \left[1 + \epsilon A_{\Lambda}^{\mathrm{sink}}(q, R_{0}) P_{1}(x)\right] + \mathcal{O}(\epsilon^{2}) \, ,
\end{equation}
where
\begin{equation}
A_{\Lambda}^{\mathrm{sink}}(q, R_{0})
=
h(q) \left[
1 +
\frac{
\Phi_{\alpha}^{(0)} \mathcal{J}_{\alpha}(q)
}{
\Phi_{\alpha}^{(1)}
- k_{\alpha}(q) \Phi_{\alpha}^{(0)}
}
\right] \, .
\label{eq:A_lambda_sink_endmatter}
\end{equation}

The limiting forms follow directly.
In the full-cone limit,
\begin{equation}
\lim_{q \to \infty} A_{\Lambda}^{\mathrm{sink}} = 1 \, .
\end{equation}
For finite $\Lambda$, the empty-cone limit reduces to the common absorbing-boundary problem and gives
\begin{equation}
\lim_{q \to 0} A_{\Lambda}^{\mathrm{sink}}
=
\frac{\sqrt{\Lambda}}{2}
\coth{\left[
\frac{\sqrt{\Lambda}}{2}
\ln{\left(\frac{1}{R_{0}}\right)}
\right]} \, ,
\label{eq:A_lambda_sink_empty_endmatter}
\end{equation}
which is identical to the phase-resolved result in Eq.~\ref{eq:A_lambda_phase_empty} and leads to Eq.~\ref{eq:A_sink_limits} in the main text.
We conclude that despite the different loss prescriptions, the empty- and full-cone limits of the continuous loss and phase-resolved calculations are identical.

\end{document}